`
\documentclass[12pt,prd,showpacs,amsmath,amssymb]{revtex4}

\usepackage{graphicx}
\usepackage{dcolumn}
\usepackage{bm}

\usepackage{amsmath}
\usepackage{graphicx}
\usepackage{amssymb}

\DeclareMathOperator{\plog}{plog}
\DeclareMathOperator{\Ai}{Ai}
\renewcommand{\vec}[1]{{\boldsymbol{#1}}}
\begin{document}
\begin{titlepage}

\title{Splitting The Gluon?}
\author{Antti J. Niemi}
\email{Antti.Niemi@teorfys.uu.se}
\homepage{http://www.teorfys.uu.se/people/antti}
\affiliation{Department of Theoretical Physics,
Uppsala University,
P.O. Box 803, S-75108, Uppsala, Sweden}
\affiliation{Yukawa Institute for Theoretical Physics, Kyoto University, 
Kyoto 606-8502, Japan}
\author{Niels R. Walet}
\email{Niels.Walet@manchester.ac.uk}
\affiliation{School of Physics, The University of Manchester, 
P.O. Box 88, Manchester M60 1QD, United Kingdom}

\date{\today}
 
\begin{abstract}
In the strongly correlated environment of high-temperature cuprate
superconductors, the spin and charge degrees of freedom of an electron
seem to separate from each other. A similar phenomenon may be present
in the strong coupling phase of Yang-Mills theories, where a
separation between the color charge and the spin of a gluon could play
a role in a mass gap formation. Here we study the phase structure of a
decomposed $SU(2)$ Yang-Mills theory in a mean field approximation, by
inspecting quantum fluctuations in the condensate which is formed by
the color charge component of the gluon field. Our results suggest
that the decomposed theory has an involved phase structure. In
particular, there appears to be a phase which is quite reminiscent of
the superconducting phase in cuprates. We also find evidence that
this phase is separated from the asymptotically free theory
by an intermediate pseudogap phase.
\end{abstract}
\maketitle
\end{titlepage}

\section{Introduction}

There seem to be some remarkable similarities between high-temperature
cuprate superconductivity in condensed matter physics and the problem
of mass gap in the Yang-Mills theory of particle physics. It
appears that in both cases the basic theoretical problem is the same,
the absence of a natural condensate to describe the symmetry breaking
that takes place.  In high-temperature superconductors electrons do
not form conventional Cooper pairs and the standard BCS-description of
superconductivity can not be applied in any obvious manner. There is
no obvious alternative choice of condensate that leads to
superconductivity. In a very similar way, in the case of Yang-Mills
theories we do not have any natural candidate for a condensate of the
correct dimension, that describes the mass gap of gluons. Could it
then be that in both cases the condensate has a similar origin?

It is definitely worth some effort to try and apply similar techniques
to both problems. One promising method in the context of high
temperature superconductivity is the slave-boson description, which
has been studied actively \cite{and1,bask,lee}. This approach is based
on the curious idea that, in the strongly correlated environment of
cuprate superconductors, the electron (or hole) is no longer a
fundamental mode of excitation, and thus electronic modes do not
behave like a structureless fundamental object. Instead the electron
can be interpreted as a composite particle, constructed from two
quasi-particles.  One of these is described by a charge neutral,
spin-$1/2$ fermionic operator $f_{i\sigma}$ where $i$ is the site
label and $\sigma = \uparrow, \downarrow$ is the spin index. This
operator corresponds to a particle called a spinon, and it carries the
(statistical) spin degree of freedom of the electron. The other
excitation is described by a spinless bosonic operator $b^T_i =
(b_{i1}, b_{i2})$. It corresponds to a particle which is called a
holon and it carries the electric charge of the electron. In terms of
these two operators, the electron operator $c_{i\sigma}$ decomposes as
\begin{equation}
c_{i\sigma} = \frac{1}{\sqrt{2}} b^\dagger_i \psi_{i\sigma}\quad,
\label{comp}
\end{equation}
where we have combined the spinon operators as
\begin{equation}
\psi^T_{i\sigma} = (f_{i\sigma}, \epsilon_{\sigma \tilde\sigma} 
f^\dagger_{i\tilde\sigma})\quad.
\end{equation}
The decomposition (\ref{comp}) also introduces an internal $U(1)$
gauge symmetry, since it is invariant under the simultaneous
change-of-phase transformation
\begin{equation}
b_i \to e^{i\theta} b_i, \qquad \psi_{i\sigma} \to
e^{i\theta} \psi_{i\sigma}.
\label{compact1}
\end{equation}
As a result we have a compact $U(1)$ gauge interaction between the
spinon and holon. Under normal circumstances we expect that the
strength of this $U(1)$ interaction increases with increasing energy,
to the effect that at high energies the spinon and holon are confined
into a (point-like) electron. But in a strongly correlated environment,
such as in a cuprate superconductor, the spin and the charge of the
electron can become independent excitations \cite{and1,bask,lee}.
This leads to a rather involved phase diagram, with several different
regions \cite{lee}. One of the easiest ways to study the phase
structure is using a mean-field theory. This is obtained by
integrating over the fermions $\psi_{i\sigma}$, and one finds that
($d$-wave) superconductivity occurs when the remaining bosonic holon
field $b_{i}$ condenses,
\begin{equation}
\langle b_i^\dagger b_i \rangle = \Delta_b \neq 0.
\label{conden1}
\end{equation}
Of substantial interest is also the possibility that the system can
enter a pseudogap phase. This is a precursor to the superconducting
phase with the characteristic property that even though the underlying
symmetry is broken, the effective bosonic order parameter $\Delta_b$
vanishes due to quantum fluctuations.

Curiously, a very similar picture seems to emerge in the case of a
pure four dimensional $SU(2)$ Yang-Mills theory.  In analogy to the
slave-boson decomposition of an electron, the off-diagonal components
of the non-abelian gluon field become composite particles, with a
separation between their color-charge and spin degrees of freedom
\cite{ludvig1} (see also \cite{oma1,ludvig2}).  
Here we shall study the phase structure of
the decomposed gauge theory, by following the mean-field approach to
high-temperature superconductivity. We first construct a mean-field
state where we integrate over the charge neutral spin degree of
freedom of the off-diagonal gluon. We propose that in the strong
coupling regime the spinless color-charge carrier of the gluon becomes
condensed. The ensuing phase is analogous to the superconducting phase
in cuprates. Furthermore, in
analogy with cuprate superconductors we also find evidence that there
is an intermediate pseudogap phase, a cross-over
region between the superconducting-like phase and the asymptotically-free
deconfined limit of the Yang-Mills theory. 

\section{Slave-Boson Decomposition In Yang-Mills}

The slave-boson decomposition of the $SU(2)$
gauge field $A^a_\mu$ ($a=1,2,3$ and $\mu=0,1,2,3$) 
proceeds as follows 
\cite{ludvig1,oma1}: We first separate the diagonal 
Cartan component $A^3_\mu = A_\mu$ from the off-diagonal 
components $A^{1,2}_\mu$, and combine the latter 
into the complex field $W_\mu = A^1_\mu + i A^2_\mu$. We then 
introduce a complex vector field $\vec{ e}_\mu$ with
\[
\vec{ e}_\mu \vec{ e}_\mu = 0 \qquad \text{and}
\qquad \vec{ e}_\mu \vec{ e}^*_\mu = 1.
\]
We also introduce two spinless
complex scalar fields $\psi_1$ and $\psi_2$. 
The ensuing decomposition of $W_\mu$ is \cite{ludvig1}
\begin{equation}
W_\mu = A^1_\mu + i A^2_\mu = \psi_1 \vec{ e}_\mu + 
\psi_2^* \vec{ e}^*_\mu.
\label{dec1}
\end{equation} 
This is clearly a direct analogue of Eq.~(\ref{comp}),  
a decomposition of $W_\mu$ into spinless bosonic 
scalars $\psi_{1,2}$ which describe the gluonic holons 
that carry the color charge of the $W_\mu$, 
and a color-neutral spin-one vector $\vec{ e}_\mu$ which is 
the gluonic spinon that carries the statistical spin 
degrees of freedom of $W_\mu$. 

In general, the present gluonic slave-boson
decomposition is not gauge invariant. But in a proper
gauge it can be given a gauge invariant meaning and in
particular the combination
\begin{equation}
\rho^2 = \rho_1^2 + \rho_2^2 = \langle |\psi_1|^2 \rangle 
+ \langle |\psi_2|^2\rangle
\label{zakh}
\end{equation} 
of the gluonic holons becomes a gauge invariant quantity.
For this we introduce \cite{zakharov,oma1}
\begin{eqnarray}
\int \rho^2 = \int ( \rho_1^2 + \rho_2^2) 
= \int \left[(A_\mu^1)^2 + (A_\mu^2)^2\right]
= \int W_\mu W_\mu^* 
\label{gf}.
\end{eqnarray}
This is in general gauge dependent. But if we consider the
gauge orbit extrema of (\ref{gf}) with respect to 
the full $SU(2)$ gauge transformations, these extrema
are by construction gauge independent quantities. Moreover, 
the gauge orbit extrema of (\ref{gf}) correspond to field 
configurations $W_\mu$ which are subject to a
background version of the maximal abelian gauge \cite{oma1},
\begin{equation}
(\partial_\mu + ig A_\mu) W_\mu  =  0,
\label{mag}
\end{equation}
which is widely used in lattice studies \cite{hay}. 
In the sequel we shall assume that the gauge fixing 
condition (\ref{mag}) has been implemented. The
slave-boson decomposition then acquires a gauge invariant
meaning, and in particular the condensate 
(\ref{zakh}) is a gauge invariant quantity.

As in (\ref{compact1}), the decomposition (\ref{dec1}) remains intact
when we change phases according to
\begin{equation}
\psi_{1,2} \to e^{i\theta} \psi_{1,2} \quad\text{and}\quad
\vec{ e}_{\mu} \to e^{-i\theta} \vec{ e}_\mu
\label{intu1}.
\end{equation}
This determines an internal compact $U(1)$ gauge structure. A compact
$U(1)$ gauge theory is known to be confining when the coupling is
sufficiently strong \cite{polyakov}. 
The confining phase is separated by a first order
phase transition from the deconfined weak coupling phase. 
Furthermore, since the running of the  
$\beta$-function of the compact $U(1)$ leads to an increase of the
coupling with increasing energy, we expect that at high energy the
gluonic holon and spinon become confined by an increasingly strong
compact $U(1)$ interaction to the effect that the high energy
Yang-Mills theory describes asymptotically free and pointlike gluons,
as it should.

But at low energy and in a strongly correlated environment,
maybe in the interior of hadronic particles, the internal $U(1)$
gauge interaction (\ref{intu1}) can become weak
and the spin and the color-charge degrees
of freedom of the gluon can separate from each other. 
If in analogy with (\ref{conden1})
the spinless color-carriers then condense
\[
\rho^2 = \rho_1^2 + \rho_2^2 =
\langle \psi_1^\dagger \psi_1 \rangle   + 
\langle \psi_2^\dagger \psi_2  \rangle  
= \Delta_\psi \not=0\quad,
\]
we have a mass gap and the theory is in a phase which is
very similar to the holon condensation phase of cuprate
superconductors. 

In the case of high-temperature superconductivity the basic criterion
for the validity of the slave-boson decomposition is a dynamic one:
The decomposition can occur only if the ensuing Hamiltonian admits a
natural interpretation in terms of the decomposed variables. In particular,
in the relevant background the holon and spinon operators 
should indeed describe proper particle states. We
propose that the same criterion can also be adopted to Yang-Mills
theories. A decomposition of the gauge field $A^a_\mu$ in terms of
other fields leads to a valid description of the phase structure, only
if the decomposed action has a natural structure and particle
interpretation in terms of the new
variables. In the case of Eq.~(\ref{dec1}) this criterion turns out to
be satisfied. If we write the Yang-Mills action in terms of the
decomposed variables, it admits a natural interpretation as a two-gap
abelian Higgs model \cite{oma1}. This suggests that the present
Yang-Mills version of the slave-boson decomposition might actually
identify the correct dynamical degrees of freedom that
describe the non-perturbative phases of the theory.

\section{The Mean-Field Theory}

In the case of cuprate superconductors, the phase structure 
can be investigated using a mean-field theory that emerges
when the original theory is averaged over the 
electronic spinon field. We now proceed in an analogous manner, 
and average the $SU(2)$ Yang-Mills action both over the color-spinon 
$\vec{ e}_\mu$ and the Cartan component $A_\mu$ of the gauge field. 
Since we are only interested in the phase structure of the ensuing 
mean-field theory, it is sufficient to consider the free energy in 
a London limit where the slave-boson condensates
\[
\rho_{1,2}^2 = \langle |\psi_{1,2}|^2 \rangle 
\]
are spatially uniform.

The integration over $A_\mu$ and $\vec{ e}_\mu$ can be performed
in various different ways. Our starting point is the 
one-loop result of Ref.~\cite{lisa}, which yields for the (London limit)
condensates the dimensionally transmutated free energy
\begin{equation}
F = \frac{1}{8} g^2 (\rho_1^2 - \rho_2^2)^2 \cdot \biggl( 1+
\frac{22}{3}\frac{g^2}{(4\pi)^2} \cdot \biggl[ \ln \frac{|\rho_1^2 -
\rho_2^2|}{\Lambda^2} - \frac{25}{6} \biggr] \biggr)\quad.
\label{lisa1}
\end{equation}
Here $\Lambda$ is the renormalization scale 
and $g^2$ is the ($\Lambda$ dependent) coupling constant and
a finite renormalization $\Lambda \to \bar\Lambda$ sends
$g \to \bar g$ with the familiar relation
\begin{equation}
\bar g^2 = \frac{g^2}{1+ \frac{22}{3}\frac{g^2}{(4\pi)^2}
\ln (\bar \Lambda/\Lambda) }\quad,
\label{scal1}
\end{equation}
or in infinitesimal form
\[
\Lambda \cdot \frac{d g}{d\Lambda} = 
\beta(g) = - \frac{22}{3}\frac{g^3}{(4\pi)^2}\quad.
\]

The minima of (\ref{lisa1}) 
are highly nondegenerate, and located on the 
$\rho_1 > 0$, $\rho_2 >0$ branch of the hyperbola
\begin{equation}
|\rho_1^2 - \rho_2^2|_{\text{min}} = 
\Lambda^2 \exp \biggl( -\frac{24\pi^2}{11g^2} +
\frac{11}{3}\biggr)\quad.
\label{minval1}
\end{equation}
Along these hyperbola, the value of the free energy is
\begin{equation}
F_{\text{min}} = - \frac{11}{24}\frac{g^4}
{16\pi^2} \Lambda^4 \exp\biggl(
-\frac{48\pi^2}{11g^2} + \frac{22}{3}\biggr)\quad.
\label{minval2}
\end{equation}
 
Since (\ref{lisa1}) is a one-loop approximation it can not be used for
providing numerically accurate predictions. For this, high-precision
Monte Carlo simulations are needed. But if one is only interested in
the qualitative features of the phase diagram, the explicit form
(\ref{lisa1}) is adequate, as shown below.

\begin{figure}
\begin{center}
\includegraphics[width=10cm,keepaspectratio,clip]{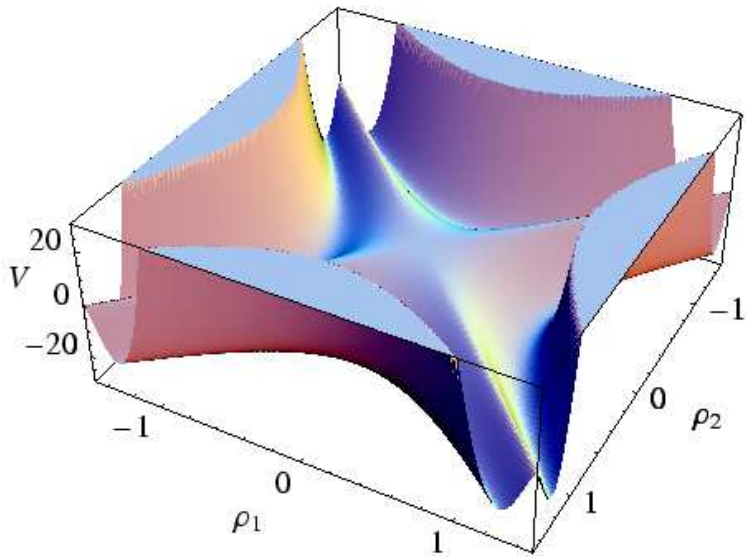}\end{center}
\caption{\label{fig:pote} The free energy $F$ in equation (\ref{lisa2}), for
$\lambda=100$. The physical branch corresponds to the quadrant
$\rho_{1,2}\geq 0$. [color online.]
}
\end{figure}

Indeed, if 
we assume that the Yang-Mills $\beta$-function has no zeroes so that
\[
\int \frac{dx}{\beta(x)} < 0\quad,
\]
the minimum values (\ref{minval1}) and (\ref{minval2}) can be
represented 
in the renormalization group invariant form
\[
(\rho_1^2 - \rho_2^2)^2_{\text{min}} =
\Lambda^4 \exp \biggl( - 4 \int\limits^g \frac{dx}{\beta(x)} + \frac{22}{3}
\biggr),
\]
and
\[
F_{\text{min}} = - 4 \Lambda^4 \cdot \int\limits^g \beta(x)dx \cdot 
\exp\biggl( -4 \int\limits^g 
\frac{dx} {\beta (x)} + \frac{22}{3} \biggr) = -4 \int\limits^g \beta(x) dx 
\cdot (\rho_1^2 - \rho_2^2)^2_{\text{min}}.
\]
Consequently we expect that the qualitative features of our conclusions
have a validity which extends beyond the one-loop level. 
For the present purposes it is sufficient to 
start from the notationally simpler version
\begin{equation}
\tilde F = \frac{1}{2} (\rho_1^2 - \rho_2^2)^2 \cdot \biggl( 1 + \tilde
\lambda \cdot \biggl[ \ln \frac{|\rho_1^2-\rho_2^2|}{\Lambda^2} - 
\alpha \biggr] \biggr)\quad.
\label{lisa1b}
\end{equation}
We normalize $F$ with the factor 
$(1-\lambda \alpha)$, set $\Lambda^2 = 1$ and redefine
\[
\lambda = \frac{1}{2}\frac{\tilde \lambda}{1-\tilde \lambda \alpha}
 \quad, \]
and arrive at the final version of the free
energy that we shall use in our analysis: 
\begin{equation}
F = \frac{1}{2}(\rho_1^2 - \rho_2^2)^2 \cdot \biggl( 1 + \lambda \cdot
\ln (\rho_1^2-\rho_2^2)^2 \biggr) \quad.
\label{lisa2}
\end{equation}
In figure~\ref{fig:pote} we have plotted this free energy for
$\lambda=100$, on the entire $(\rho_1,\rho_2)$ plane. 
The generic features of this potential, a ridge along
the lines $\rho_1=\pm\rho_2$, and a narrow hyperbolic valley on both
sides of these lines, are independent of $\lambda$, but the depth of the
valleys and steepness of the potential are more prominent for 
larger values of $\lambda$, as used here.

The Landau pole at 
\begin{equation}
\tilde \lambda = 1/\alpha \label{eq:landaupole}
\end{equation}
separates the strong coupling region with $\tilde \lambda > 1/\alpha$
from the weak coupling region with $\tilde \lambda < 1/\alpha$. Since
the latter region includes the small coupling limit of the original
Yang-Mills theory and since the free energy (\ref{lisa1}) can only be
reliable for weak coupling, we shall in the following concentrate on
the region $ 0 \leq \tilde \lambda \leq {1}/{\alpha}.  $ Notice that
in terms of the redefined coupling $\lambda$ in (\ref{lisa2}), this
corresponds to the region of a positive $\lambda$. In particular, if
$\alpha$ in (\ref{lisa1b}) is large, the strong coupling limit $\lambda
\to \infty$ in (\ref{lisa2}), does not necessarily correspond to a strong
coupling limit of the original model (\ref{lisa1b}).

\section{Classical Aspects}

We first consider the properties of the free energy
(\ref{lisa2}) at a classical level, where we do not include the
quantum fluctuations in the spatially uniform London
limit condensates $\rho_{1,2}$. This free energy
has the following classical scaling symmetry \cite{martin},
\begin{eqnarray}
\rho' & = & \frac{\rho}{c} \quad,\nonumber \\
\lambda' & = & \frac{\lambda}{1+2\lambda\ln(c^{2})}
\quad,\nonumber \\
F' &=& \frac{1}{c^4} \cdot \frac{F}{1+2\lambda \ln c^2}  
\quad.\label{scaleall}
\end{eqnarray}
We can employ this scaling symmetry to restore
the parameter $\Lambda$ in the free energy; see (\ref{lisa1b}),
(\ref{lisa2}). Indeed, it is obvious that this scaling symmetry 
reflects the renormalization
group symmetry of the original Yang-Mills theory, with
the scaling transformation of the coupling constant
$\lambda$ a version of (\ref{scal1}). 

In addition, as a function on the entire $(\rho_1, \rho_2)$
plane the free energy has a 
discrete symmetry since it only depends
depends on a polynomial
combination of the condensates
\[
F(\rho_1,\rho_2)=f\left((\rho_1^2 - \rho_2^2)^2\right).
\]
and in addition we also 
have the gauge invariant polynomial
combination in Eq.~(\ref{gf}),
\begin{equation}
\rho^2 = \rho_1^2 + \rho_2^2\quad.
\label{gf2}
\end{equation}
We are interested in linear transformations that act on the $(\rho_1,
\rho_2)$ plane and leave both polynomials intact. These polynomials
are exactly the basic invariants that generate the octonic dihedral
group $D_4$ (also called $4mm$), which is the nonabelian symmetry
group of the square. 

The four branches of the hyperbola that minimize (\ref{lisa2}),
\begin{equation}
(\rho_{1}^{2}-\rho_{2}^{2})^2 = \exp\left(\frac{1-\lambda}
{\lambda}\right)\quad,
\label{grhyp}
\end{equation}
are separated by (non-analytic) ridges along the lines 
$\rho_{1}=\pm \rho_{2}$, and mapped to
each other by the $D_4$ transformations.
At the minima along the hyperbolic valleys 
the free energy is given by
\begin{equation}
E_{\text{min}} =- \frac{1}{2}\exp\left(\frac{1-\lambda}{\lambda}\right)\quad.
\label{grene}
\end{equation}
This ground state is highly degenerate, but
the combination on the left-hand side  of (\ref{grhyp}) is 
not the proper gauge invariant condensate. The gauge invariant 
condensate is given by Eq.~(\ref{gf2}), and we can employ it
to remove the infinite degeneracy of the hyperbolic vacuum:

From (\ref{grhyp}) we conclude that
the ground state value $\rho^2 = v^2$ of the gauge invariant
condensate (\ref{gf2}) is bounded from below by a non-vanishing
quantity,
\begin{equation}
\rho^2 = \rho_1^2 + \rho_2^2 = v^2  \geq  |\rho_1^2 - \rho_2^2| =
\exp\left(\frac{1-\lambda}{2\lambda}\right).
\label{lower1}
\end{equation}
When $v^2$ is larger than the lower bound in (\ref{lower1}),
there are eight solutions $(\rho_1,\rho_2)$ to the equations 
that define the vacuum
\begin{eqnarray}
\rho_1^2 + \rho_2^2 & = & v^2 \quad,\nonumber \\
\rho_1^2 - \rho_2^2 & = & \pm \exp\left( \frac{1-\lambda}{2\lambda}\right)
\quad.\label{vaceq1}
\end{eqnarray}
But when $v^2$ coincides with the lower bound there are only four 
solutions,
\begin{eqnarray}
\rho_1=\pm v \quad&\& &\quad \rho_2=0 
\quad,\nonumber \\
\rho_1=0 \quad&\& &\quad \rho_2=\pm v
\quad.\label{vaceq2}
\end{eqnarray}
which correspond to the vertices of the hyperbola.
The solutions are mapped onto each other by the dihedral group
$D_4$, and selecting any one as the ground state breaks the $D_4$
symmetry. 

The solutions of (\ref{vaceq1}) describe the generic situation
where both condensates are non-vanishing. The solutions are
$D_4$-degenerate, but we remove this degeneracy when we select
the (physical) $\rho_{1,2}\geq 0$ quadrant. The remaining ground state
is doubly degenerate under exchange of $\rho_1$ and $\rho_2$, which
correspond to the physical scenario that in general the London
limit densities are unequal.

Finally, the degenerate solutions (\ref{vaceq2}) correspond to 
the limit where one of the two condensates vanishes, and again by 
selecting the physical quadrant $\rho_{1,2} \geq 0$ we remove the 
degeneracy.

According to (\ref{lower1}) the ground state value of (\ref{gf2}) is
non-vanishing for all non-vanishing values of the coupling constant
$\lambda $. This suggests that in the Yang-Mills theory the gauge
invariant condensate (\ref{zakh}) is also nonvanishing for all
non-vanishing values of the coupling constant. This would mean that
the mass gap in the Yang-Mills theory is present for all values of the
coupling, and it vanishes only asymptotically in the short distance
limit where the gluons become asymptotically free and massless. 

\section{ Quantum Mechanics - Numerical Approach}

The classical treatment of the mean-field theory in the previous
section suggests that the condensate (\ref{zakh}) is always
non-vanishing, hence a mass gap is present for all
nontrivial values of the coupling. 
We now want to inspect what effects spatially
homogeneous quantum fluctuations around the classical mean-field value
have on this condensate.  For this we need to improve the free energy
so that it also includes the contribution from the momenta $\pi_{1,2}$
that are canonically conjugate to the (spatially homogeneous)
condensates $\rho_{1,2}$. For computational simplicity we 
consider these condensates to be defined over the entire 
$(\rho_{1},\rho_{2})$ plane. This results in a $D_4$ symmetry, and 
by selecting the physically relevant values $\rho_{1,2} \geq 0$ 
for the condensates we then break this discrete symmetry.

The conjugate momenta are the generators
of spatially homogeneous translations. Their inertia is undefined, 
and we therefore add a parameter $M$.  
The improved free energy can be interpreted as a Hamiltonian
\begin{equation}
H=\frac{1}{2M}\left(\pi_{1}^{2}+\pi_{2}^{2}\right)+
\frac{1}{2}\left(\rho_{1}^{2}-\rho_{2}^{2}\right)^2
\left(1+\lambda\ln\left(\rho_{1}^{2}-\rho_{2}^{2}\right)^{2}
\right)\quad.
\label{Ham}
\end{equation}
It corresponds to the effective action
\begin{equation}
S_{\text{eff}} = \int\limits_0^T dt \bigl(
\pi_1 \dot \rho_1 + \pi_2 \dot \rho_2 - H[\pi,\rho] \bigr)\quad,
\label{effect}
\end{equation}
and the equations of motion for $S_{\text{eff}}$ 
are invariant under the following extension \cite{martin}
of the scaling transformation (\ref{scaleall})
\begin{eqnarray}
\rho' & = & \frac{\rho}{c} \quad,\nonumber \\
\pi' & = & \sqrt{\frac{1}{1+2\lambda\ln(c^{2})}}\hskip 0.6mm
\frac{\pi}{c^2}\quad,\nonumber \\
\lambda' & = & \frac{\lambda}{1+2\lambda\ln(c^{2})}
\quad,\nonumber \\
t' & = & \sqrt{1+2\lambda\ln(c^{2})}
\cdot ct \quad.\label{scaleall2}
\end{eqnarray}
We also note that the action (\ref{effect}) is clearly invariant under the
dihedral $D_4$ symmetry group.

In order to study the effects of quantum fluctuations in the
condensates we investigate the solutions of Schr\"odinger equation
\begin{equation}
-\frac{\hbar^{2}}{2M} \left(\partial_{1}^{2} + \partial_{2}^{2}
\right) \psi(\rho_{1},\rho_{2}) + \frac{1}{2}
\left(\rho_{1}^{2}-\rho_{2}^{2}\right)^2 \left(1+\lambda\ln
\left(\rho_{1}^{2}-\rho_{2}^{2}\right)^{2}\right) \psi(\rho_{1},\rho_{2})
=E\psi(\rho_{1},\rho_{2}).
\label{Schr}
\end{equation}
We have studied this Schr\"odinger equation (\ref{Schr}) numerically,
using a highly-accurate finite difference approximation, on grids of
varying size and spacing, using up to $400\times400$ grid points.  We
have analyzed both the ground-state wave function and several of the
low-lying excited-state wave functions when the coupling constant
$\lambda$ in (\ref{Schr}) varies for fixed $M$.  According to the
relation between (\ref{lisa1b}) and (\ref{lisa2}), this surveys the
phase structure of the theory at couplings below the Landau pole.

We note that since the Schr\"odinger equation (\ref{Schr}) is
invariant under the action of the dihedral $D_4$, the wave functions
can be chosen to have definite $D_4$ transformation
properties. Unfortunately, most discussions of point groups, see
e.g.~\cite{Butler}, look for representations in 3D space, where one
can distinguish between the groups $C_{4v}$ and $D_4$, but these
groups act identically in the $xy$ plane. 
Following Mulliken's ($A,B$ 1D irrep; $E$
2D irrep) or Koster's notation ($\Gamma_i$)
as discussed in Ref.~\cite{Butler}, we have five possible
representations of this group in two dimensions, see table \ref{tab:group}.

\begin{table}
\caption{Character table (trace over classes of elements of 
the representation matrices) for $D_4$. 
The five classes on the top line are, respectively: $E$ identity;
$2C_{4z}$ rotations over $\pm\pi/2$ around the $z$-axis; $C_{2z}$ a
rotation over $\pi$ around the $z$-axis; $2\sigma_y$ reflections in the
$x$ or $y$-axis; $2\sigma_{xy}$ reflections in the lines 
$x=\pm y$.} \label{tab:group}
$\displaystyle
\begin{array}{lrrrrrlll}
& E & 2C_{4z} &  C_{2z} & 2 \sigma_y & 2 \sigma_{xy} 
& \text{representative wave function}\\
\hline
A_1=\Gamma_1 & 1 & 1       &  1      & 1          
& 1            & f(x^{2}+y^{2})\\
A_2=\Gamma_2 & 1 & 1       &  1      & -1         
& -1         & xy(x^2-y^2) \\
B_1=\Gamma_3 & 1 & -1      & 1       & 1          
& -1         & x^2-y^2 \\
B_2=\Gamma_4 & 1 & -1      & 1       & -1         
& 1          & xy \\
E=\Gamma_5   & 2 &  0      & -2      &  0         
& 0          & (x,y)\text{ and } (x+y,x-y)\\
\end{array}
$
\end{table}

In that table we have also listed representative wave functions for
all the irreps.  Looking at the symmetry of the wave functions, we
expect $A_1$, $B_1$ and one of the $E$ cases with zeroes on the lines
$x=\pm y$ to form four almost degenerate states as $\lambda$ grows
large, as borne out by figure~\ref{fig:l400} below.

When $\lambda \to 0$ the Schr\"odinger equation (\ref{Schr}) reduces
to the $x^2y^2$ model which has been studied in detail in
\cite{simon}. In particular, it has been established that the spectrum
of the $x^2y^2$ model is discrete, the eigenstates are normalizable,
and the ground state energy is separated from $E=0$ by a non-vanishing
gap.
\begin{figure}
\begin{center}\includegraphics[width=15cm, keepaspectratio,clip]{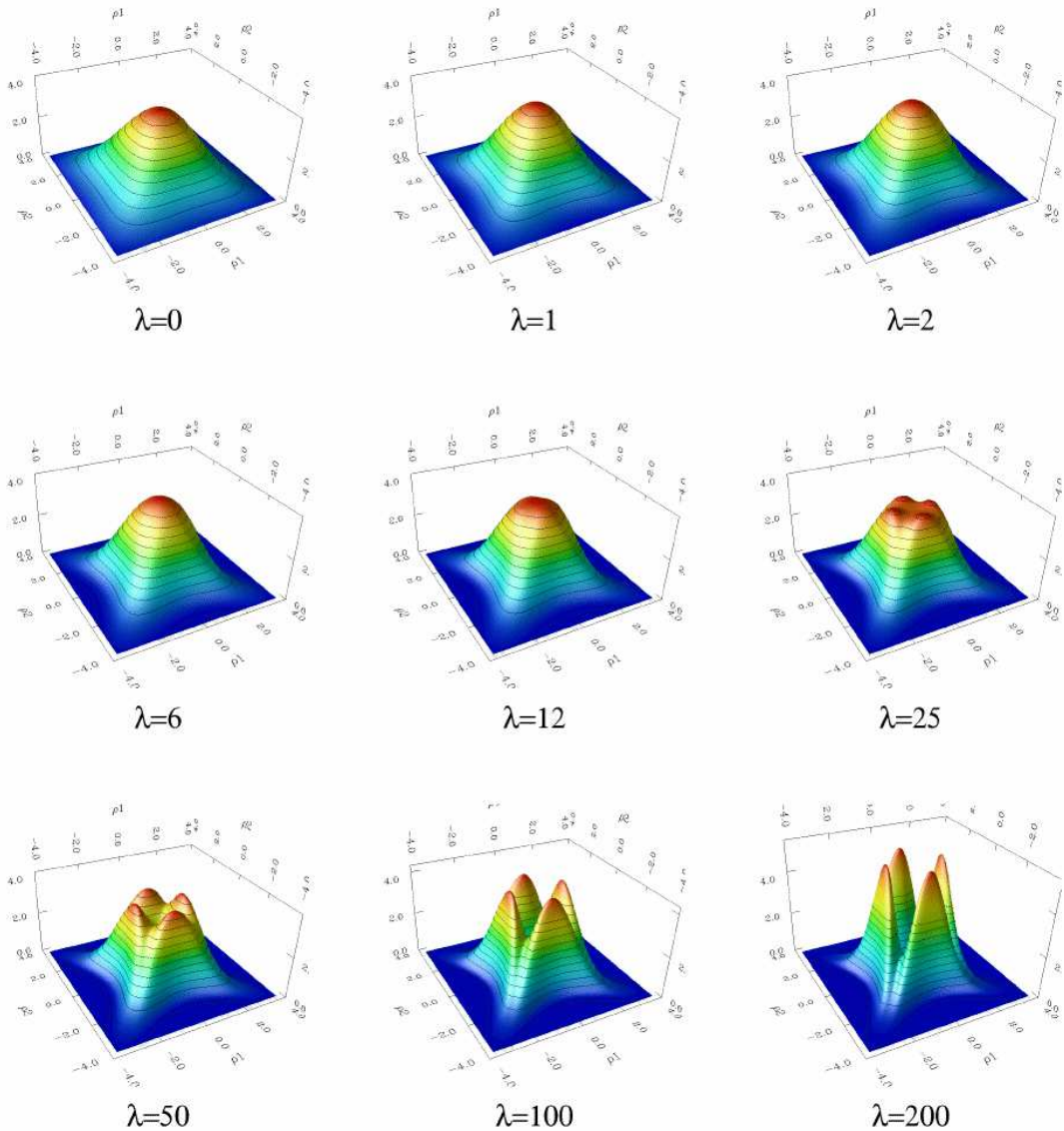}
\end{center}
\caption{\label{fig:wfs}The ground state wave function for 
$\lambda$ from $0$ to $200$. A result for a larger value of 
$\lambda$ is given in 
figure~\ref{fig:l400}. [color online.]
}
\end{figure}

In figure \ref{fig:wfs} we depict the behavior of the numerically
constructed ground state wave function for different values of
$\lambda$ for $M=1$. Very similar behavior is found for other values
of $M$, but as analyzed in more detail below, the similarity is
greatest if we compare solutions for identical values of $\lambda M$.
We find that the wave function exhibits three different kinds of
qualitative behavior.  There is a weak coupling region $0 < \lambda M
< \lambda_a \approx 10$, an intermediate coupling region $\lambda_a <
\lambda M < \lambda_b \approx 500$ and a strong coupling region
$\lambda_b < \lambda$. In all cases the ground-state wave function
lies in the lowest symmetric representation ($A_1$) of $D_4$. These
regions have the following characteristic features:

\begin{figure}
\begin{center}\
\includegraphics[width=10cm, keepaspectratio,clip]{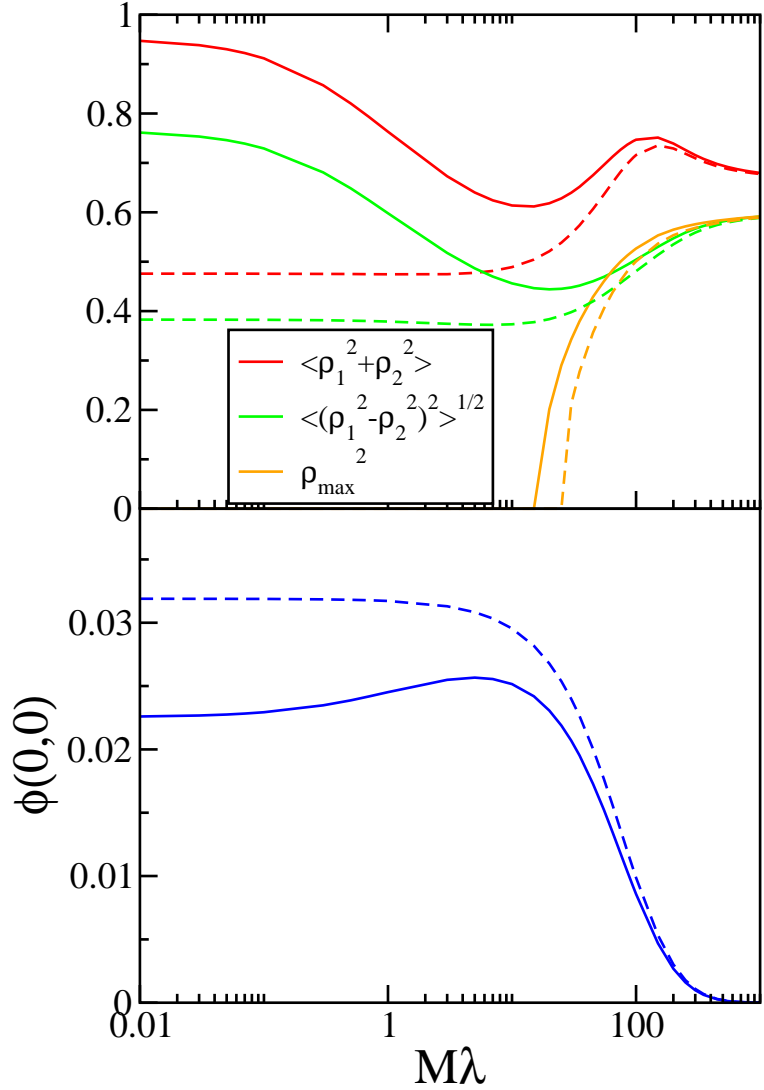}
\end{center}
\caption{\label{fig:subtract}
The upper panel shows the two condensates $\rho^2$ and $((\rho_1^2 -
\rho_2^2)^2)^{1/2}$ , and the distance of the maximum value of the ground state
wave function to the origin.  The lower panel gives the 
value of the ground state wave function at
the origin.  The solid lines show the values for $M=1$, and the dashed
lines for the larger value $M=8$. [color online.]}
\end{figure}

\paragraph{Weak coupling:}
In the weak coupling region with $M \lambda < \lambda_a \approx 10$,
we find that the ground state wave function is qualitatively 
reminiscent of the
ground state wave function in the $x^2y^2$ model, in particular
it has a single
maximum which is located at the origin of the $(\rho_1 , \rho_2)$
plane.  We also find that the value of the wave function at its
maximum varies very slowly as a function of $\lambda$, especially for
a large value of $M$, which means a more tightly localized wave
function, see figure (\ref{fig:subtract}); Both the shape of the
ground state wave function and the location of its maximum suggest,
that in this weak coupling region quantum fluctuations tend to
restore the system towards the symmetric state $\rho^2 \approx 0$ so
that there would not be any mass gap in the underlying Yang-Mills theory. 

\begin{figure}
\begin{center}\
\includegraphics[width=10cm, keepaspectratio,clip]{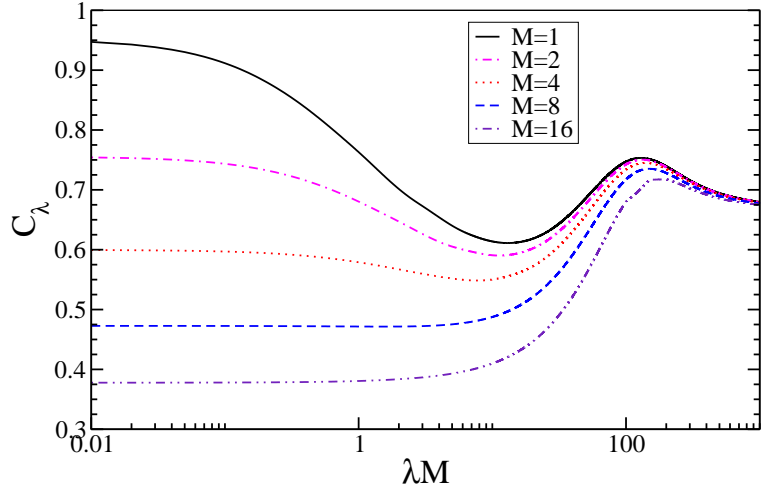}
\end{center}
\caption{\label{fig:compare}
The condensate (\ref{zakh}), 
as a function of $M\lambda$ for various values of $M$, as indicated in the
plot. [color online.]
}
\end{figure}

Clearly, we find non-zero condensate values for any finite value 
of $M$. As suggested by figure~\ref{fig:compare}, the condensate 
goes to zero at $\lambda=0$, as $M$ goes to infinity but there 
remains a cross-over to a broken phase at stronger coupling. 
This is consistent with the fact, that in the limit of vanishing 
coupling the Yang-Mills theory describes free massless gluons. 

Independent of $M$, when $\lambda $ approaches $\lambda_a$ 
the value of $\rho_\lambda$
overshoots the classical value given by the right-hand side of
(\ref{grhyp}). Consequently the expectation value (\ref{zakh})
detects the presence of symmetry breaking and the ensuing
nontriviality of the condensate, even though this is not reflected in
the location of the maximum value of the ground state wave function.

Such a behavior where the condensate (\ref{zakh}) detects a
symmetry breaking while the wave function tends to retain the symmetry, is
reminiscent of the pseudogap phase \cite{lee}. As a consequence we
propose that in the weak coupling region $0 < \lambda < \lambda_a$
the underlying Yang-Mills theory is in a pseudogap phase,
a cross-over region which terminates in the asymptotically 
free theory as $\lambda \to 0$. Presumably this region is intimately
related to a Coulomb-like phase in the Yang-Mills theory.

\paragraph{Intermediate coupling:}
When $10 \approx M\lambda_a < M\lambda < M\lambda_b \approx 500$ 
there is a clear
qualitative change in the behavior of both the ground state wave
function and the condensates (\ref{zakh}). For these values of the coupling
the origin in the $(\rho_1,\rho_2)$ plane becomes a local minimum, 
and instead there are now four maxima in the wave function. These maxima
are all located the same distance $\rho_{\text{max}}$ from the origin,
and related to each other by the $D_4$ symmetry. Both the value of the
ground state wave function at the origin, and the condensate (\ref{zakh})
decrease essentially linearly in the logarithmic scales of figure
\ref{fig:subtract}, while the value of $\rho_{\text{max}}$ very
rapidly approaches the classical limiting value determined by
(\ref{grhyp}), as $\lambda \to \lambda_b$. 
In particular, as $\lambda \to \lambda_b$ the value of
the condensate (\ref{zakh}) becomes less than its classical bound
in (\ref{grhyp}), but is still clearly bounded from below.

In this intermediate coupling region both the ground state wave
function and the condensate (\ref{zakh}) behave similarly, and in a
manner which suggests that the underlying Yang-Mills theory has a mass
gap. Indeed, the behavior is quite reminiscent of the superconducting
phase in cuprate superconductors.  We find it natural to propose that
this region of the coupling constant describes a superconducting
mass-gap phase of the Yang-Mills theory, maybe a magnetic dual to
the confinement phase.

\paragraph{Strong coupling:}
When $\lambda \to \lambda_b \approx 1000$ we detect a new transition,
towards a strong coupling regime $\lambda_b < \lambda$. Now the
value of the ground state wave function essentially vanishes at the
origin, see figure \ref{fig:subtract}. The value of the condensate
(\ref{zakh}) again increases, and asymptotically approaches the
value $\rho_{\text{max}} = \exp(-1/4)$ which is the classical
$\lambda \to \infty$ lower bound value (\ref{lower1})
for the minimum distance between the potential minimum and the 
origin. Indeed, for the entire strong coupling 
region $\lambda_b < \lambda$ we find that the difference between
the classical and quantum values of the condensate is very small,  
suggesting that in this region one of the condensates
essentially vanishes. Consequently as $\lambda \to \infty$ 
the system becomes driven towards a degenerate ground state where 
one of the condensates asymptotically vanishes, while the 
other becomes asymptotically determined by the classical theory.

The strong coupling region retains the major characteristics of the
intermediate coupling region: There is a mass gap, and the ground
state wave function is peaked at a nontrivial value of the
condensate, even though in the infinite coupling limit one of 
the (quantum) condensates seems  to vanishes asymptotically--
it seems that the two quantities  $\rho^2$ and $((\rho_1^2 -
\rho_2^2)^2)^{1/2}$ coincide as $\lambda \rightarrow \infty$.
But in this region the ground state wave function has the additional
characteristic property that it (essentially) vanishes in a
neighborhood around the origin, thus becoming (essentially) separated
into four disjoint components. This means that the lowest four states,
consisting of two one-dimensional representations and one
two-dimensional one, become degenerate. While we do recognize that in
a finite dimensional quantum mechanical model there always remains a
(vanishingly small) tail of the wave function at the origin, the
numerically observed vanishing of the wave function in the vicinity of
the origin is very definite. Consequently we envision, that in the
underlying field theory with its infinite number of degrees of
freedom, there is a true transition where the tunneling between the
four dihedrally symmetric branches of the ground state wave function
becomes totally suppressed.

Finally, we find that the at these high values of $\lambda$
the wave functions of the four lowest states become degenerate,
see figures \ref{fig:l400}. The symmetries of these states correspond
to $A_1$, $E$ and $B_1$ irreps, as argued above.
\begin{figure}
\begin{center}
\includegraphics[ width=10cm, keepaspectratio,clip]{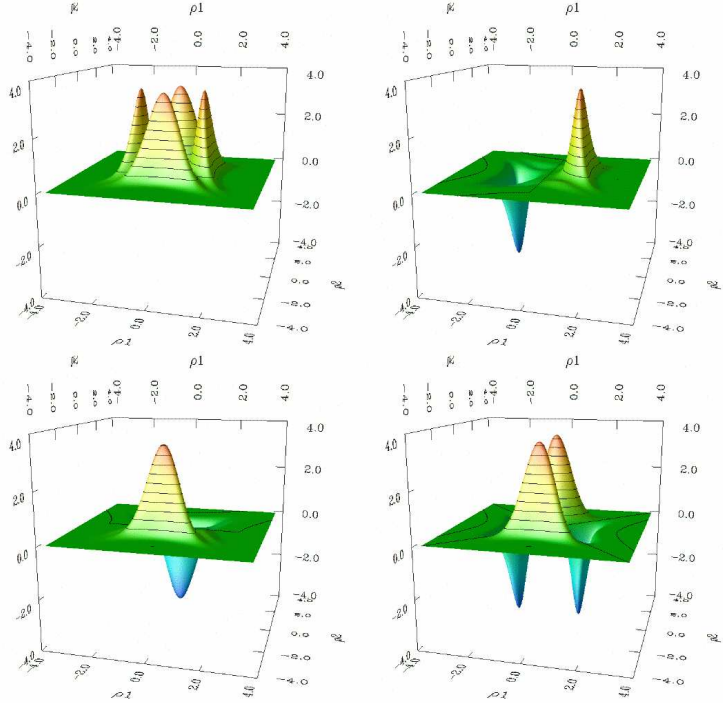}
\end{center}

\caption{The four lowest eigenstates (from left to right and top to bottom) of
the problem (\ref{Schr}) for $\lambda=400$. \label{fig:l400}[color online.]}
\end{figure}

\section{Quantum Mechanics - Asymptotic Analysis}

The present model is a generalization of the $x^2y^2$ model, a notoriously
complex system \cite{simon}. While the $\lambda \to 0$ limit 
of our numerical 
results reproduce the known properties of the $x^2y^2$ model,
there is a need to confirm the main features of our $\lambda \not=0$ 
results by formal analysis. For this, we now consider the
relevant asymptotic behavior of the ground state wave function. 

\subsection{Large distance behavior of the wave function}
For $\lambda =0$ the behavior of the solutions of (\ref{Schr}) are
known and have been discussed in detail in \cite{simon}: The spectrum
is discrete, the eigenstates are normalizable, and the ground state
energy is separated from $E=0$ by a non-vanishing gap. Our numerical
investigations suggest that these conclusions persist for
non-vanishing values of $\lambda$. We now proceed to verify this using
asymptotic analysis of the Schr\"odinger equation (\ref{Schr}). In
particular, we wish to establish that the wave function is indeed
normalizable.

When $\rho_1 \not= \pm\rho_2$ the potential in (\ref{Schr}) is bounded
from below by a positive quadratic form. Consequently any peculiar,
unexpected behavior in the ground state wave function must be
concentrated near the lines where $\rho_{1}^{2}=\rho_{2}^{2}$. This is
best studied in hyperbolic coordinates \cite{Moon}
\begin{equation}
\xi=\frac{1}{2}(\rho_1^{2}-\rho_2^{2}), \quad\text{and}\quad\eta=
\rho_1 \rho_2 \quad.\label{eq:hyperbolic}
\end{equation}
We note that even though these coordinates only cover half 
of the $(\rho_1, \rho_2)$ plane, we can use them to study 
the full behavior of the wave function for large values of
$\rho_1 \approx  \pm \rho_2$. In these coordinates 
the Schr\"odinger operator is
\[
\hat{H}=-\sqrt{\xi^{2}+\eta^{2}}\left(\partial_{\xi\xi}+
\partial_{\eta\eta}\right)+2\xi^{2}(1+\lambda\ln(4\xi^{2}))
\]
and we are particularly interested in the behavior of the wave
function for large values of $\eta$ and small values of $\xi$.

We first consider the known $\lambda=0$ case, this leads us to
the Schr\"odinger equation 
\[
-|\eta|(\phi_{\xi\xi}+\phi_{\eta\eta})+2\xi^{2}\phi=E\phi\quad.
\]
We wish to implement an \emph{asymptotic} separation of variables.
For this we select $\eta$ to be large and positive (alternatively
large and negative). Since the potential depends on $\xi$ alone, 
we can introduce a transformation in this variable that 
allows separation to (almost) take place: 
We write
\[
x=\xi/\eta^{1/4} \quad \& \quad y=\eta^{1/2}
\]
and
\[
\phi(\xi,\eta)=f(x)g(y)
\] 
With this Ansatz we get
\begin{multline}
yg(y)\left(-f''(x)+2x^{2}f(x)\right)-\frac{f(x)g''(y)}{4}
+\frac{1}{y}\left[-\frac{5xg(y)f'(x)}{16}+\frac{f(x)g'(y)}{4}+
\frac{xf'(x)g'(y)}{4}\right]\\
-\frac{x^{2}g(y)f''(x)}{16y^{2}}=Ef(x)g(y)\quad.
\label{sequ}
\end{multline}
Thus, the problem is separable to leading order in $1/y$,
\begin{eqnarray}
-f''(x)+2x^2 f(x)=\tau f(x)\\
-\frac{g''(y)}{4}+\tau yg(y)=0
\end{eqnarray}
The function $f$ is clearly one of the harmonic oscillator 
states, and for the ground state of our Schr\"odinger equation 
we must have the lowest energy eigenstate of the harmonic 
oscillator. The equation for $g$
then becomes
\[
-\frac{g''(y)}{4}+\sqrt{2}yg(y)=0
\]
since for large values of $y$ the value of the
energy becomes irrelevant. 
For a normalizable wave function, the only acceptable 
solution is
\begin{equation}
g(y)=\Ai(2^{5/6}y)
\end{equation}
which decays rapidly. This is consistent with our
numerical simulations, and the ($\lambda = 0$) results 
in \cite{simon}: The ``tendrils'' of the wave function 
along the potential valleys are indeed decaying very rapidly.
In original coordinates,
\begin{equation}
\phi(\rho_1,\rho_2) \approx \exp(-\frac{1}{8}
(\rho_1^2-\rho_2^2)^2/(\rho_1\rho_2))\Ai(2^{5/6}(\rho_1\rho_2)^2)
\end{equation}

We now consider the general case: As above, our approach is based on
asymptotic separation of variables obtained by rescaling $\xi$, which
allows us to combine the terms multiplying $f''$ with the rescaled
potential. Let us therefore look at the $\xi
\to \alpha \xi$ scaling of the general potential, as studied in
Eq.~(\ref{scaleall}),
\begin{equation}
V_{\lambda}(\xi)  =2\xi^{2}(1+\lambda\ln4\xi^{2}) \quad.
\label{asyscale1}
\end{equation}
We find
\begin{equation}
V_{\lambda}(\alpha\xi) =\alpha^{2}(1+\lambda\ln\alpha^{2})2\xi^{2}
\left(1+\frac{\lambda}{(1+\alpha\lambda)}\ln4\xi^{2}\right)
=\beta V_{\tilde{\lambda}}(\xi)\quad,
\label{asyscale2}
\end{equation}
where
\begin{equation}
\beta= \alpha^{2}(1+\lambda\ln\alpha^{2}) \quad\&\quad
\tilde{\lambda} =\frac{\lambda}{(1+\alpha\lambda)}\quad.
\label{asyscale3}
\end{equation}
If we now make $\alpha$ dependent on $\eta$, and define
$x=\xi/\alpha(\eta)$, we get a matching condition by requiring a
common $\eta$-dependent factor for the leading second derivative with
respect to $x$ and the potential
\[
\eta\alpha^{2}=\alpha^{2}(1+\lambda\ln\alpha^{2})
\]
which has the solution
\begin{equation}
\alpha=\left(\frac{2\eta}{\lambda\plog\left(\frac{2\eta e^{2/\lambda}}
{\lambda}\right)}\right)^{1/4}\quad,
\end{equation}
where $\plog$ is the {}``product logarithm'' (inverse to $xe^{x}$). 

\begin{figure}
\begin{center}\includegraphics[ width=7cm, keepaspectratio,clip]{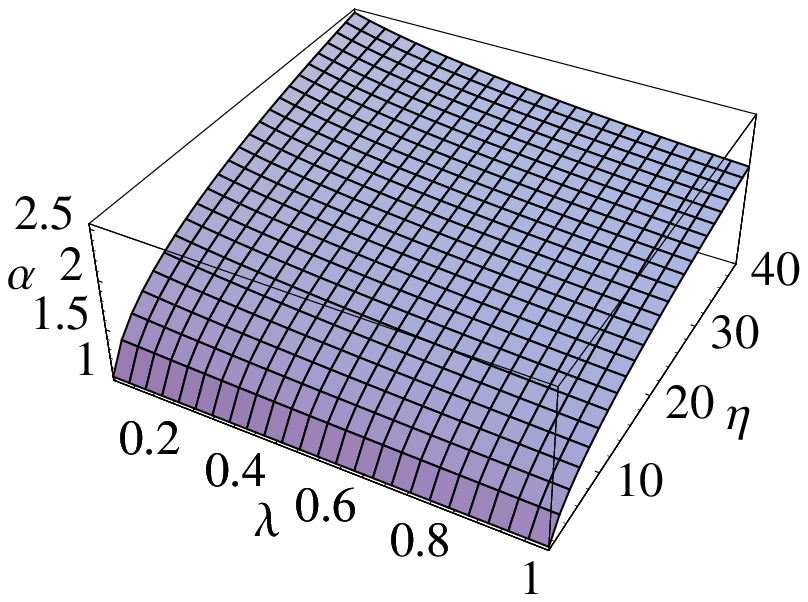}%
\includegraphics[width=7cm,keepaspectratio,clip]{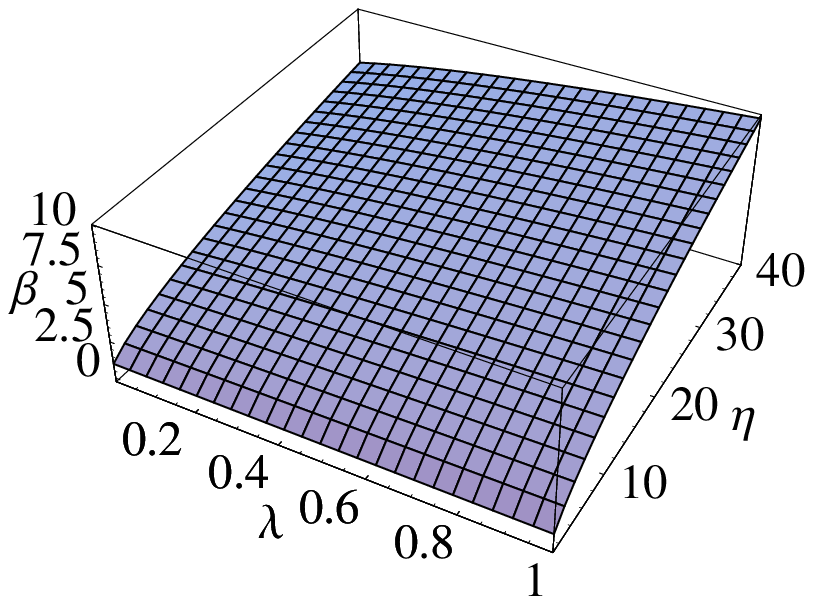}\end{center}
\caption{\label{fig:alphabeta}The functions $\alpha$ (left) and $\beta$
(right) vs. $\lambda$ and $\mu$.[color online.]}
\end{figure}

\begin{figure}
\begin{center}\includegraphics[ width=6cm,clip]{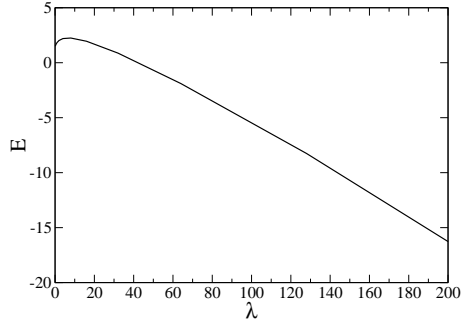}\end{center}
\caption{\label{fig:Lowestev}
The lowest eigenvalue of the 1D Schr\"odinger equation (\ref{eq:fl})
as a function of $\lambda$.
}
\end{figure}

The functions $\alpha(\eta)$ and $\beta(\eta)$ (see Fig.~\ref{fig:alphabeta})
are increasing functions of $\eta$, and will provide us with a second
expansion parameter. Now we separate variable in $x$ and $\eta$ as before,
\[
\phi(\xi,\eta)=f(x)g(\eta)\quad,
\]
which leads us to the following generalization of (\ref{sequ})
\begin{multline}
-\eta g''(\eta)f(x)+\beta(\eta)g(\eta)\left(-f''(x)+V_{\tilde{\lambda}
(\eta)}(x)f(x)\right)- \\
\frac{\beta(\eta)^{2}\alpha'(\eta)^{2}}{4\eta}
g(\eta)\left(3xf'(x)+x^{2}f''(x)\right) \\
-\beta(\eta)\alpha'(\eta)xf'(x)g'(\eta)+\frac{1}{2}\beta(\eta)
\alpha''(\eta)xf'(x)g(\eta)=Ef(x)g(\eta)\quad.
\label{sequ2}
\end{multline}
We now ignore all but the first three terms-- it can
be verified numerically that all other terms are small--and separate variables
\begin{eqnarray}
-f''(x)+V_{\tilde{\lambda}}f(x)&=&
\epsilon(\tilde{\lambda}(\eta)) f(x)\label{eq:fl}\quad,\\
-\eta g''(\eta)+\beta(\eta)\epsilon(\tilde{\lambda}(\eta))
g(\eta)&=&0\quad.
\end{eqnarray}
The value $\beta(\eta)$ is larger than $\eta^{1/2}$,
its value when $\lambda=0$, as can be seen in Fig.~ \ref{fig:alphabeta}.
The eigenvalue $\epsilon(\tilde{\lambda})$ is larger than 
$\sqrt{2}$, if we take $\eta$ large enough so that $\tilde \lambda={\cal O}(1)$, see Fig.~\ref{fig:Lowestev}.
Since $\tilde{\lambda}\rightarrow 0$ for $\eta\rightarrow\infty$,
we then have a rapid decay of the wave function for large $\eta$,
as expected.

\begin{figure}
\begin{center}\includegraphics[width=8cm,clip]{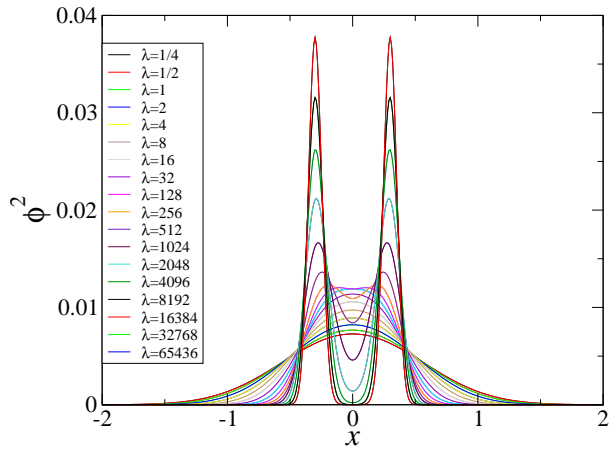}\end{center}
\caption{\label{fig:Lowestf2}
The lowest eigenfunction of  (\ref{eq:fl}) as a function of $\lambda$. [color online.]}
\end{figure}

The equation for $f(x)$ is interesting for other reasons as well;
as shown in figure~\ref{fig:Lowestf2}, we find that for large values 
of $\lambda$ the wave function separates into two parts. Remembering
that  $x=0$ corresponds to the lines $\rho_1=\pm\rho_2$,
this supports our assertion that the wave function separates
in 4 disjoint parts for large $\lambda$.

\subsection{Large values $\lambda>\lambda_b$}
In the region of large coupling, the hyperbolic valleys of the free
energy become very deep. We are interested in the asymptotics of the
ground state wave function, when it becomes separated into four
disjoint components.  We continue to utilize the hyperbolic
coordinates, but we shall now expand around the minima of the free
energy.

In hyperbolic coordinates, the minimum of the free energy
\[
\xi_{0} =\pm\frac{1}{2}\exp\left(-\frac{1+\lambda}{2\lambda}\right)
\]
remains very close to $\xi=0$
(less than $\frac{1}{2\sqrt{e}}=0.303\ldots$ which is the 
value for $\lambda\rightarrow\infty$). Consequently the 
previous asymptotic analysis remains valid, and we can
immediately conclude that the wave functions are decaying 
rapidly.

However, it is also of interest to consider the limit of a very deep
potential directly, and for this we introduce coordinates from the minimum,
scaled with $\xi_0$,
\[ \xi = \xi_{0}(x\pm1) \quad\&\quad
\eta = \xi_{0} y\quad.
\]
We then expand in powers of $x$, keeping leading terms only.
The Hamiltonian simplifies to
\begin{eqnarray}
H &=&-\frac{1}{\xi_{0} M}\sqrt{(1+x)^{2}+y^{2}}(\partial_{xx}
+\partial_{yy})+2\lambda\xi_{0}^{2}(x+1)^{2}\left[\log(x+1)^{2}-1\right]
\nonumber\\
&\approx&-\frac{1}{\xi_{0}M}\sqrt{(1+y^{2}}(\partial_{xx}
+\partial_{yy})+2\lambda\xi_{0}^{2}(-1+2x^{2})\quad.
\end{eqnarray}
With $\mu=\lambda\xi_{0}^{3}$ and $\epsilon=E\xi_{0}+\mu$
we this leads to the eigenvalue problem
\begin{equation}
-\frac{\sqrt{1+y^{2}}}{M}(\partial_{xx}\phi+\partial_{yy}\phi)+4\mu x^{2}
\phi=\epsilon\phi\quad.
\end{equation}
We now wish to consider the properties of solutions to 
this Schr\"odinger equation: We substitute 
\[
\phi(x,y)=f(z)g(y)\quad,
\] 
with 
\[
z=x(1+y^{2})^{1/8}\quad.
\] 
When we ignore terms containing lower order or mixed derivatives in addition
of powers of $x$, we find that the equation takes the form
\[
(1+y^{2})^{1/4}g(\frac{1}{M}\partial_{zz}f-4\mu z^{2}f)+f\sqrt{1+y^{2}}
\frac{1}{M}\partial_{yy}g=\epsilon fg\quad.
\]
Assuming again the lowest harmonic oscillator eigenstate for $f$, 
\[
f=\exp(-(z/b_{z})^{2}/2)\quad,
\]
with 
\[
b_{z}=(2\mu M)^{-1/4}\quad,
\] 
we find for $g$,
\[
-\frac{\sqrt{1+y^{2}}}{M}\partial_{yy}g+\sqrt{2\mu}(1+y^{2})^{1/4}g=\epsilon g\quad.
\]
In order to obtain an analytic solution we shall assume $\mu$ to
be so large that we again can make the harmonic approximation.
This gives
\[
-\partial_{yy}g+\sqrt{\mu/2}y^{2}g=(\epsilon-\sqrt{2\mu})g\quad.
\]
Thus 
\[
g=\exp(-(y/b_{y})^{2}/2)\quad,
\]
with $b_{y}=(8/M \mu)^{1/8}$, and the ground state energy is given
by 
\[
\epsilon=(2\mu/M)^{1/2}+(\mu/(2 M)^3)^{1/4}\quad.
\]
The energy for the original problem can thus be expressed as
\[
E=\frac{(2\mu/M)^{1/2}+(\mu/(2M)^3)^{1/4}-\mu}{\xi_{0}}\quad.
\]
which is a good approximation only when the wave function $fg$ has no
overlap with those from the remaining three valleys - and when our
harmonic approximations are valid. 

We also conclude that the scaling of the condensates observed in the
previous section is indeed taking place; the width of both $f$ and $g$
depends on the combination $\mu M=\lambda M \xi_0^3$, and for large
$\lambda$ $\xi_0$ is approximately constant, leading to the observed 
scaling in $\lambda M$. Since the wave function contract slowly to the 
maximum point (most slowly for $g$), we find that indeed we
have in the limit $\lambda \rightarrow \infty$ the case where
one of the two condensates disappears, as stated above. A further
 numerical analysis using the separated wave function
confirms that this approach is very slow, and we must go to extremely
high values of $\lambda$ to see the point where we can't distinguish between
the maximum and the expectation value of $\rho^2$.

\section{Conclusions}

In conclusion, we have investigated the phase structure of pure
$SU(2)$ Yang-Mills theory using a slave-boson
decomposition of the gauge field. We have employed a 
mean-field approximation where we account only for spatially 
homogeneous fluctuations in the gluonic holon fields. Our 
analysis suggests that the decomposed theory has an involved 
phase diagram, resembling that of cuprate superconductors.
At intermediate couplings, there seems to be a gapped
phase which is separated from the asymptotically free high 
energy limit by a pseudogap phase. 
Furthermore, we find that a mass gap appears to
persists in the strong coupling limit even though asymptotically 
one of the two
holon condensates appears to vanishes.

In our analysis we have employed a version of the
maximal abelian gauge. In this gauge we have the advantage, 
that many 
results are available from first principle lattice 
simulations; see \cite{hay} for a review. 
In particular, it has been observed \cite{hay} that 
the (electric) confinement of color relates to
the condensation of magnetic monopoles in the
dual Higgs phase.
Here we have inspected a (renormalization group invariant)
perturbative one-loop approximation to the Yang-Mills
effective action, in terms of decomposed variables that
have a natural magnetic interpretation. Our results do
not account for topologically nontrivial configurations,
consequently it is not directly clear how our results could
relate to the (electric) color confinement as observed
in the lattice simulations. For a comparison, we need 
a first principles numerical lattice analysis in 
terms of the separated spin and charge variables. We also need
a better understanding of electric-magnetic
duality in terms of these variables. 
However, even at the level of the (crude) approximation
that we employ here, it appears that when formulated
in terms of the separate spin and charge variables the
Yang-Mills theory has a very rich phase structure,
not easily described in terms of the conventional
gluonic variables. Our results suggests that the 
possibility of a spin-charge separation in the Yang-Mills theory
may occur, and deserves to be addressed
by extensive first-principles lattice
simulations. Furthermore, there is a need to address
theoretical issues such as electric-magnetic duality and the 
description of the Yang-Mills theory in
terms of the spin-charge separated dual variables. 

Indeed, if gluons can become decomposed 
into their independent holon and spinon 
components, it could have deep consequences to our understanding
of the fundamental structure of matter.

\acknowledgments
AJN has been supported by a grant from VR (Vetenskapsr\.adet) and by a
STINT Thunberg Fellowship; The work of NRW has been supported by a
grant from the ESPRC and as part of the PPARC SPG ``Classical Lattice
Field Theory''.  We thank M. Chernodub, B. DeWit, J. Hoppe,
M. Polikarpov and Z.Y. Weng for discussions.

\end{document}